\documentclass[aps,preprint,showpacs]{revtex4}

\begin{document}

\title{Spatiotemporal discrete multicolor solitons}
\author{Zhiyong Xu, Yaroslav V. Kartashov$^{\dag}$, Lucian-Cornel Crasovan$^{*}$,
Dumitru Mihalache$^{*}$, and Lluis Torner}

\affiliation{ICFO-Institut de Ciencies Fotoniques, and Department
of Signal Theory and Communications, Universitat Politecnica de
Catalunya, 08034 Barcelona, Spain}

\begin{abstract}
We have found various families of two-dimensional spatiotemporal
solitons in quadratically nonlinear waveguide arrays. The families
of unstaggered odd, even and twisted stationary solutions are
thoroughly characterized and their stability against perturbations
is investigated. We show that the twisted and even solutions
display instability, while most of the odd solitons show
remarkable stability upon evolution.
\end{abstract}


\maketitle

\affiliation{ICFO-Institut de Ciencies Fotoniques, and Department of Signal
Theory and Communications, Universitat Politecnica de Catalunya, 08034
Barcelona, Spain}

\section{Introduction}

Since their first experimental observation \cite{Torruellas},
quadratic solitons have been demonstrated in a variety of
materials and geometries. Spatial, temporal, and spatiotemporal
solitons in quadratic media have been extensively investigated
both experimentally and theoretically (for detailed reviews, see
\cite{steg1,Etrich,Buryak,Lluis}). Quadratic solitons also exist
in the form of discrete entities, namely strongly localized wave
packets forming in nonlinear waveguide arrays. Since their
theoretical prediction in 1988 in cubic nonlinear media
\cite{christodoulides}, discrete optical solitons have attracted a
steadily growing interest because of their potential applications
in switching and routing devices \cite
{christodoulides1,Lederer,ole1}. The discrete solitons that form
in tight-coupled waveguide arrays made of quadratic nonlinear
media have been comprehensively investigated
\cite{Sukhorukov,Bang,Bang1998,Peschel,Kobyakov1,Malomed} due to
the rich variety of effects that are possible with them. Such
richness may be further enhanced by combining the features of
both, continuous and discrete soliton families present in
spatiotemporal discrete solitons, a possibility that we address
here.

In the last two decades the concept of optical spatiotemporal solitons
(STSs), referred as light bullets in the three-dimensional case \cite
{Silberberg}, has been attracting attention as a unique opportunity to
create a self-supporting fully localized object. The existence of STS's in
quadratic nonlinear materials was theoretically predicted \cite{STchi2} and
thereafter experimentally realized in a two-dimensional geometry involving
one temporal and one spatial coordinate \cite{Liu}. The existence and
properties of continuous-discrete spatiotemporal solitons has been
extensively investigated in cubic nonlinear media and stable odd solitons
have been shown to exist \cite{Aceves1,Aceves2,Aceves3,Laedke1,Laedke2}. It
was shown that the cubic weakly-coupled waveguide arrays act as collapse
compressors \cite{Aceves1,Aceves2,Aceves3}. In contrast with the cubic
spatiotemporal solitons, the quadratic ones do not display collapse in both
two- and three-dimensional geometries \cite{Berge}. A still open problem,
not analyzed so far, is the existence of space-time solitons in nonlinear
waveguides with quadratic nonlinearity, that is, the existence of discrete
spatiotemporal multicolor solitons.

In this paper we investigate in detail the existence and stability of three
representative families of two-dimensional spatiotemporal solitons in
quadratic nonlinear waveguide arrays. We assume, in addition to the temporal
dispersion of the pulse, the contribution of the discrete diffraction, that
arises because of the weak coupling between neighboring waveguides.

Discrete soliton solutions were classified as {\it staggered} and
{\it unstaggered} ones (see, for example, Ref. \cite{Falk}). The
staggered solutions display out-of-phase fields between the
neighbor noncentral waveguides whereas the unstaggered ones
display in-phase fields in these noncentral waveguides. Inside
each of these classes of solitons (staggered and unstaggered) one
can find solutions with different topologies, dictated mainly by
the energy and phase distribution in the central waveguides. Thus,
one can have (i): {\it odd} solitons, for which most part of the
energy is located in one central waveguide and the energy
distribution across the waveguide array is symmetric with respect
to this central waveguide, (ii): {\it even} solitons, for which
most part of the energy is equally distributed in the two central
waveguides, the fields in these central waveguides being in-phase
and of equal amplitudes, and (iii): {\it twisted} solitons, for
which most part of the energy is equally distributed in the two
central waveguides, but the fields in the two central waveguides
are out-of-phase.

Here we will restrict ourselves to three representative families
of continuous-discrete unstaggered solitons, namely the odd
soliton (see Fig. 1(a)), the even soliton (see Fig. 1(c)) and the
twisted soliton (see Fig. 1(d)). Note that for the twisted
soliton, the fundamental frequency field is, in fact, an
anti-symmetric one (the $\pi $ jump of phase occurs only between
the two central waveguides), whereas the second harmonic field is
a symmetric one (having the form of an even discrete soliton). For
all the solutions we deal with, the temporal profile, i.e. the
shape of the pulses propagating in a specific waveguide, is a
bell-shaped symmetric one (see Fig. 1(b), below). Besides these
stationary solutions, there exist a whole ``zoology'' of localized
solutions, including staggered solitons, dark or dark-bright
solitons, but their study is beyond the scope of the present work.

\section{Model and Stationary Solutions}

The evolution of the spatiotemporal two-component field in
quadratic nonlinear waveguide arrays in a degenerate
second-harmonic generation geometry may be described by the
following set of nonlinearly coupled reduced differential
equations:
\begin{eqnarray}
&&i\frac{\partial u_n}{\partial \xi}=-c_u\left( u_{n-1}+u_{n+1}\right) +\frac{%
g_1}2\frac{\partial ^2u_n}{\partial \tau ^2}-u_n^{*}v_n\exp \left(
-i\beta
\xi\right),   \nonumber \\
&&i\frac{\partial v_n}{\partial \xi}=-c_v\left( v_{n-1}+v_{n+1}\right) +\frac{%
g_2}2\frac{\partial ^2v_n}{\partial \tau ^2}-u_n^2\exp \left(
i\beta \xi\right) ~,
\end{eqnarray}
where $u_n$ and $v_n$ represent the normalized amplitudes of the fundamental
frequency (FF) and second-harmonic (SH) fields in the $n$th waveguide, with $%
n=-N,...-1,0,1,...,N$, $2N+1$ being the number of waveguides,
$^{*}$ means complex conjugation, $c_{u,v}$ and $g_{1,2}$ are the
linear coupling coefficients and group-velocity dispersion (GVD)
coefficients, respectively, and $\beta $ is the wave-vector
mismatch. The evolution variable $\xi$ denotes the normalized
propagation distance along the waveguides. The dynamical system
(1) admits several conserved quantities including the energy flow
and Hamiltonian which read
\begin{equation}
I=\sum_n\int \left( \left| A_n\right| ^2+\left| B_n\right|
^2\right) d\tau\, ,
\end{equation}
\begin{eqnarray}
&&H=-\sum_n\int \left[ \frac{g_1}2\left| \frac{\partial A_n}{\partial \tau }%
\right| ^2+c_u\left( A_nA_{n+1}^{*}+A_n^{*}A_{n+1}\right) \right.
\nonumber \\
&&+\frac 12(A_n^2)^{*}B_n+\frac{g_2}4\left| \frac{\partial
B_n}{\partial \tau }\right| ^2+\frac{c_v}2\left(
B_nB_{n+1}^{*}+B_n^{*}B_{n+1}\right)
\nonumber \\
&&\left. -\frac \beta 2\left| B_n\right| ^2+\frac
12A_n^2B_n^{*}\right] d\tau\, ,
\end{eqnarray}
where we have defined $A_n\equiv u_n$, and $B_n\equiv v_n\exp \left( -i\beta
\xi\right) $. The stationary solutions of Eqs. (1) have the form $%
u_n=U_n\left( \tau \right) \exp \left( ib_1\xi\right) $ and
$v_n=V_n\left( \tau \right) \exp \left( ib_2\xi\right) $, where
$U_n\left( \tau \right) $ and $V_n\left( \tau \right) $ are real
functions, and $b_{1,2}$ are real propagation constants verifying
$b_2=2b_1+\beta $. Continuous-discrete solitons arise from a
balance between discrete diffraction, dispersion and quadratic
nonlinearity. The families of odd, even, and twisted stationary
continuous-discrete solitons have been obtained numerically by a
standard
relaxation method. For given coupling strengths $c_{u,v}$, dispersions $%
g_{1,2}$ and wave-vector mismatch $\beta $, the solitons families
are parametrized by the nonlinear wavenumber shift $b_1$. The
coupling coefficients $c_{u,v}$ were considered positive, and
equal, so further we introduce single parameter $C$ to describe
coupling between neighboring guiding sites. Throughout this paper
we will always consider anomalous dispersions at both frequencies and we fixed $%
g_1=-0.25$ and $g_2=-0.5$. Note that in the continuous case,
long-lived soliton-like propagation when the GVD is slightly
normal at SH is known to occur, \cite{tornerPTL,IsaacPRL} thus a
similar behavior might occur in the continuous-discrete
spatiotemporal case analyzed here.

In Fig. 2(a,b) we show the dependencies of the peak amplitude
$A_{u}$ and the temporal full width of half maximum of the pulse
in the central waveguide $W_{u}$  as a function of the coupling
coefficient $C$ for a fixed wavenumber $b_1$, and at phase matching ($%
\beta=0$). Note that,with increase of coupling strength amplitude
of odd and even solitons monotonically decreases and their width
increases, whereas the amplitude and width of the twisted solitons
are nonmonotonic function of $C$. This is illustrated also in Fig.
3 where profiles of odd solitons $|U_n\left( \tau \right)|$ at two
different coupling constants are shown. Note that with increase of
coupling constant soliton covers more guiding sites, while at
$C\longrightarrow0$ it is located primarily in the central guiding
site.

Similar to the two-dimensional (continuous-continuous) solitons in
uniform media, there exist cut-off $b_{co}$ of the nonlinear
wavenumber shift
$b_1$ depending on the sign and absolute value of the mismatch parameter $%
\beta$. Moreover, as we have an additional degree of freedom,
namely the discrete spatial coordinate, we have investigated the
dependence of the cut-off wavenumber $b_{co}$ on the coupling
coefficient $C$ for a given wave-vector mismatch. For a
phase-matched geometry ($\beta=0$), we have obtained almost linear
dependencies of the cut-off wavenumber on the coupling
coefficients for all three families of solutions we deal with (see
Fig. 2(c)). Note that cut-off for odd and even solitons are equal.
As a general rule, the stronger is the coupling the larger is the
cut-off wavenumber $b_{co}$. When $C=0$ we got $b_{co}=0$, thus
recovering the known result for the continuous quadratic solitons: $%
b_{co}=$max$\{-\beta/2,0\}$.

We also have investigated the peak amplitude and the temporal
width in the central waveguide for odd, even and twisted
continuous-discrete solitons as functions of the wave-vector
mismatch for fixed nonlinear wavenumber shift $b_1$ and linear
coupling coefficient $C$. The solitons that form for larger
phase-mismatches have larger amplitudes and are narrower than
those forming for smaller ones. This feature was observed for one-
and two-dimensional continuous solitons in quadratic media for
which at
phase matching the product \textit{peak-amplitude} $\times$ \textit{%
width-squared} is a constant quantity \cite{Lluis95}. In Fig. 2(d)
we plot the amplitude of the stationary odd soliton as function of
its temporal width. We see that outside phase-matching the
families of solitons exhibit a more complicated amplitude-width
relationship, similar to the case of continuous quadratic solitons
\cite{Lluis95}. The scaling properties of Eqs. (1) can be written
as:
\begin{eqnarray}
u_n=\psi {\tilde u}_{n},~v_n=\psi {\tilde v}_{n},~b_1=\psi {\tilde
b}_{1},  \nonumber
\\
\beta=\psi \tilde{\beta}, ~\tau = \tilde{\tau}/\sqrt{\psi},
~I=\psi^{3/2} \tilde{I},
\end{eqnarray}
where $\psi$ being the scaling parameter.

In Figs. 4(a)-4(f) we have represented the dependencies energy
flow $I$ - wavenumber $b_1$ (left column) and Hamiltonian $H$ -
energy flow $I$ (right column) that give us a deeper insight into
the properties of continuous-discrete soliton families. One can
see that odd solitons realize the minimum of Hamiltonian for a
given energy flow, thus they are expected to be the most robust on
propagation. The Peierls-Nabarro potential, that is the difference
between Hamiltonian of the odd soliton and that of the even one
\cite {Kivshar}, corresponding to the same energy flow, is
negative everywhere. From a geometrical point of view, this would
mean that odd solitons are stable in the entire domain of their
existence \cite{Cai}. Our numerical simulations, described in
detail in the next section, show that, indeed, this is the case
except for solitons at negative phase-mismatches that are unstable
only in a narrow region near cut-off (see Fig. 5(a)) \cite
{Pelinovsky1,Lluis95}.

\section{Stability Analysis}

A key issue concerning the soliton families we found is their
stability on propagation. In order to elucidate if the localized
continuous-discrete solitons are dynamically stable we have
performed both a linear stability analysis and direct numerical
simulations. We seek for perturbed solution of Eq. (1) in the form

\begin{eqnarray}
u_n(\tau,\xi) &=&\left[ U_n(\tau)+\mu f_n\left( \tau,\xi\right)
\right] \exp \left(
ib_1\xi\right),  \nonumber \\
v_n(\tau,\xi) &=&\left[ V_n(\tau)+\mu h_n\left( \tau,\xi\right)
\right] \exp \left[ i(2b_1+\beta) \xi\right] ,
\end{eqnarray}
Here $\mu $ is a small parameter, $U_n(\tau)$ and $V_n(\tau)$ are
the stationary solutions and $f_n\left( \tau,\xi\right) $ and
$h_n\left( \tau,\xi\right) $ are the perturbations. Then after
linearizing the evolution equations (1) we are left with a system
of linear coupled differential equations for the perturbations
(see, e.g., Ref. \cite{Soto1}):

\begin{eqnarray}
&&i\frac{\partial f_n}{\partial \xi}=-c_u\left( f_{n-1}+f_{n+1}\right) +\frac{%
g_1}2\frac{\partial ^2f_n}{\partial \tau ^2}-\left(
U_n^{*}h_n+V_nf_n^{*}\right) +b_1f_n,  \nonumber \\
&&i\frac{\partial h_n}{\partial \xi}=-c_v\left( h_{n-1}+h_{n+1}\right) +\frac{%
g_2}2\frac{\partial ^2h_n}{\partial \tau ^2}-2U_nf_n~+\left(
2b_1+\beta \right) h_n,
\end{eqnarray}

We have solved both this linear system and the nonlinear dynamical
equations (1) with a combined Fast-Fourier Transform, to deal with
the linear differential part in the temporal coordinate, and a
fourth-order Runge-Kutta method, to deal with the cross-coupling
terms. We have typically used 512 or 1024 points in the time
domain and we have considered tens of array sites (e.g., 61),
depending on the width of the solution whose stability is
investigated. The step length along the propagation coordinate was
of the order of 10$^{-3}$. The accuracy of the results was checked
by doubling the number of points in the transverse coordinate and
by halving the propagation step. As another check for the
evolution equations (1) we have verified the conservation of the
prime integrals (energy flow $I$ and Hamiltonian $H$). In order to
let the radiation to escape from the computation window we have
implemented transparent (absorbing) boundary conditions. We
multiply a flat-top function after every step of longitudinal
propagation distance, and this function has a very narrow tail
which is zero.

We have determined the dominant eigenvalue $\delta$ of the
linearized problem using the same approach as in Ref.
\cite{Soto1}. The method gives us only the dominant eigenvalue,
not the whole eigenvalue spectrum. This eigenvalue corresponds to
the most rapidly (exponentially) developing instability. The noisy
perturbation we consider at $\xi=0$ develops, during evolution, to
a localized eigenvector with a well defined symmetry, depending on
the type of the solution considered. In the cases where an
instability was detected, only real instability eigenvalues were
found. The dominant eigenvalue was calculated in the form

\begin{eqnarray}
&&Re(\delta)=\frac{\displaystyle 1}{\displaystyle
\Delta\xi}\frac{\displaystyle \sum_n\int_{-\infty}^{\infty} \left(
\left| f_n(\tau,\xi+\Delta\xi)\right| ^2+\left|
h_n(\tau,\xi+\Delta\xi)\right| ^2\right) d\tau}{\displaystyle
\sum_n\int_{-\infty}^{\infty} \left( \left| f_n(\tau,\xi)\right|
^2+\left| h_n(\tau,\xi)\right| ^2\right) d\tau},
\end{eqnarray}

This dominant eigenvalue tends to zero when one approaches the
stability region. The results we got for the growth-rate
calculations at negative phase-mismatch ($\beta=-3$) are
summarized in Fig. 5. They indicate instability for even and
twisted solitons \cite
{Sukhorukov} and a stability region for odd solitons which starts at $%
b_1^{stab} \approx 1.725$. This result is in good agreement with
the direct simulations of evolution Eq. (1). For positive
wave-vector mismatches or at phase-matching the growth rate
calculations indicate instability for even and twisted solitons
and complete stability for odd solitons.

Our calculations show that odd continuous-discrete solitons obey
the Vakhitov-Kolokolov stability criterion \cite{Vakhitov}, i.e.,
they are stable provided $dI/db_1>0,$ and unstable, otherwise. The
Vakhitov-Kolokolov criterion was shown also to hold for discrete
space-time solitons that exist in Kerr nonlinear media
\cite{Laedke1,Laedke2}. Moreover, the unstable odd cubic
continuous-discrete solitons can display collapse-type
instabilities, a reminiscent feature of the two-dimensional
stationary solutions of nonlinear Schr\" odinger equation, while
the unstable quadratic discrete space-time odd solitons do not
display this type of instability \cite{Berge}.

Let us stress that as compared to the one-dimensional discrete
twisted solitons forming in quadratic media that can be stable, in
specific parameter regions, in our case, the introduction of a
time coordinate leads to the destabilization of these solutions.
However, one of the central points of this work is that we found
families of stable odd continuous-discrete multicolor solitons. As
illustrated in Fig.6 (b), stable odd solitons can propagate for
huge distances without altering their shape and eliminating the
added random white noise during evolution. The case shown here
corresponds to negative wave-vector mismatch $\beta =-3$ but
similar stable evolution has been obtained for positive mismatches
and phase-matching geometries except for odd solitons from the
branch, where $dI/db_1<0$, which are unstable and will therefore
decay after a finite propagation distance (see Fig. 6(a)).

In addition, we also thoroughly investigated the decay scenarios
of the other two types of solitons: even and twisted. As stated
before we have not observed any stable even or twisted
continuous-discrete soliton. Fig. 7 shows possible instability
scenarios for unstaggered even and unstaggered twisted solitons.
We have found that perturbed even soliton typically tranforms into
odd one through increasing field oscillation in neighboring wave
guides (Fig. 7(a)), and perturbed twisted soliton usually splits
into two solitons which fly apart as when a repulsive force would
act between them (Fig. 7(b)). We have observed a $\pi$ phase
difference between the formed odd solitons and this could explain
the repulsive force between them. Note that during the splitting
process the resulting odd solitons are still locked in a specific
waveguide, they are being allowed to repel in time. This unique
feature comes with discreteness which does not allow the soliton
energy to escape from the waveguide where it was initially
located.

\section{Conclusion}

We have shown that stable, spatiotemporal continuous-discrete
solitons are possible in quadratic nonlinear waveguide arrays.
Families of unstaggered odd, even and twisted stationary solutions
have been found and thoroughly characterized. The linear stability
analysis is in agreement with the direct simulations indicating
that the odd continuous-discrete solitons obey the
Vakhitov-Kolokolov stability criterion. The salient point put
forward is that most of the spatiotemporal unstaggered odd
solitons are stable against perturbations. This result is
important in view of the generation of discrete solitons with
pulsed light in the context of the exploration of their potential
application to switching schemes
\cite{christodoulides1,Lederer,ole1}.

$^{\dag}$Also with the Physics Department, M. V. Lomonosov Moscow State
University, Moscow, Russia.

$^{*}$ Permanent address: Institute of Atomic Physics, Department of
Theoretical Physics, P.O. Box MG-6, Bucharest, Romania.

This work was partially supported by the Generalitat de Catalunya,
Instituci\'{o} Catalana de Recerca i Estudis Avancats (ICREA),
Barcelona, and by the Spanish Government through grant
BFM2002-2861. Zhiyong Xu's e-mail address is Xu.Zhiyong@upc.es.


\newpage

\textbf{Figure Captions}

Fig. 1. Amplitude profiles of the (a) odd, (c) even, and (d)
twisted solitons. Lines with circles show FF field, lines with
hexagons show SH field. In (b) the time slice in the central
waveguide ($n=0$) for odd soliton is shown. Even and twisted
solitons feature the similar temporal profile. Here  $C=0.1$,
$b_1=3$, and $\beta =3$.

Fig. 2. (a) Peak amplitude and (b) temporal width of FF wave in
the central waveguide for odd, even and twisted solitons versus
coupling coefficient at $b_1=3$ and $\beta =0$. (c) Wavenumber
cutoff versus coupling coefficient at $\beta =0$. The symbols
``o'', ``e'', and ``t'' stand for the odd, even, and twisted
solitons, respectively. (d) FF wave amplitude versus temporal
width in the central waveguide for odd soliton at $C=0.1$ and
different phase mismatches. Only stable branch has been plotted
for negative $\beta$.

Fig. 3. Profiles of odd solitons for (a) $C=0.5$ and (b) $C=1$ at
$b_1=3$, $\beta =0$. Only the modulus of the amplitude of the FF
wave is shown. The SH shows similar features.

Fig. 4. Energy flow versus wavenumber and Hamiltonian versus
energy flow for odd, even, and twisted solitons at three
representative values of phase mismatch and $C=0.1$. The labels
are the same as in Figs. 2.

Fig. 5. Growth rate versus wavenumber for (a) odd, (b) even, and
(c) twisted solitons at $\beta =-3$ and $C=0.1$.

Fig. 6. (a) Propagation of unstable odd soliton corresponding to
$b_1=1.65$ in the presence of small perturbation found upon linear
stability analysis. Perturbation amplitude $\mu=0.01$. (b)
Propagation of stable odd soliton at $b_1=1.735$ in the presence
of white noise with variance $\sigma^{2}_{noise}=0.01$. Only the
modulus of the amplitude of the SH wave is shown,  at different
propagation distances. Plots in left and right columns are shown
with the same scale for easier comparison. Phase mismatch $\beta
=-3$ and coupling constant $C=0.1$.

Fig. 7. Propagation of unstable even (a) and twisted (b) solitons
corresponding to $b_1=3$ in the presence of small perturbations
found upon the linear stability analysis. Perturbation amplitude
$\mu=0.01$. Only the modulus of the amplitude of the SH wave is
shown,  at different propagation distances.  Plots in left and
right columns are shown with the same scale for easier comparison.
Phase mismatch $\beta =-3$ and coupling constant $C=0.1$.

\end{document}